\documentclass[a4paper,11pt]{article}%
\usepackage[legalpaper,bookmarks=true,colorlinks=true,linkcolor=red,citecolor=red]{hyperref}
\usepackage{placeins}
\usepackage{multirow}
\usepackage{array}
\usepackage{caption}
\usepackage{amssymb}
\usepackage{graphicx}
\usepackage{amsfonts}
\usepackage{mathrsfs}
\usepackage{amsmath}
\usepackage{babel}
\usepackage{lastpage}
\usepackage[utf8]{inputenc}
\usepackage{graphicx}%
\usepackage{fancyhdr,enumerate}
\usepackage{color}
\usepackage[mathlines]{lineno}
\usepackage{lscape}
\usepackage{epsfig}
\usepackage[sort&compress]{natbib}
\usepackage{epstopdf}
\usepackage{subfigure}
\usepackage{float,adjustbox}
\usepackage{bm}
\usepackage{booktabs}
\usepackage{geometry}
\usepackage{geometry}
\begin{document}

\title{Bitcoin Price Forecasting Based on Hybrid Variational Mode Decomposition and Long Short Term Memory Network}

\author{Emmanuel Boadi$^{\textbf{1*}}$\\
	$^{\textbf{1}}$	School of Mathematical and Statistical Science, College of Sciences,\\ University of Texas Rio Grande Valley, Edinburg, Texas, USA.\\
	Corresponding author: \color{blue}emmanuel.boadi01@utrgv.edu}

		\date{}
		\maketitle
		\begin{abstract}
\noindent	
This study proposes a hybrid deep learning model for forecasting the price of Bitcoin, as the digital currency is known to exhibit frequent fluctuations. The models used are the Variational Mode Decomposition (VMD) and the Long Short-Term Memory (LSTM) network. First, VMD is used to decompose the original Bitcoin price series into Intrinsic Mode Functions (IMFs). Each IMF is then modeled using an LSTM network to capture temporal patterns more effectively. The individual forecasts from the IMFs are aggregated to produce the final prediction of the original Bitcoin Price Series. To determine the prediction power of the proposed hybrid model, a comparative analysis was conducted against the standard LSTM. The results confirmed that the hybrid VMD+LSTM model outperforms the standard LSTM across all the evaluation metrics, including RMSE, MAE and $R^{2}$ and also provides a reliable 30-day forecast.
\\

\end{abstract}




\section{Introduction}
Bitcoin is a decentralised digital currency that operates without a central bank or single administrator. It enables peer-to-peer transactions over a cryptographically secured network without the need for intermediaries \cite{sujatha2021bayesian}. These transactions are recorded on a public distributed ledger known as the blockchain \cite{niranjanamurthy2019analysis}. Bitcoin’s design supports fast, low-cost, and pseudonymous transactions, but its market value is highly volatile and influenced by various factors such as supply and demand, investor sentiment, media attention, and regulatory developments \cite{kayal2021bitcoin}.
This volatility presents challenges for accurate price forecasting, prompting the development of models capable of handling such fluctuations. Prior research has explored various approaches for Bitcoin price prediction. For example, Shah and Zhang \cite{shah2014bayesian} applied Bayesian regression, achieving an 89\% return in 50 days, while Hegazy and Mumford \cite{hegazy2016comparitive} used decision trees based on price derivatives to achieve 57.11\% directional accuracy. \\

\noindent
Traditional time series models such as ARIMA have also been applied, but their linear assumptions limit their ability to model Bitcoin’s nonlinear and nonstationary behavior. Karakoyun and Cibikdiken \cite{karakoyun2018comparison} compared ARIMA with Long Short-Term Memory (LSTM) networks, reporting a lower Mean Absolute Percentage Error (MAPE) of 1.40\% with LSTM versus 11.86\% with ARIMA.
Despite these efforts, many traditional models struggle to fully capture Bitcoin’s inherent complexity. Variational Mode Decomposition (VMD), introduced by Dragomiretskiy and Zosso \cite{dragomiretskiy2013variational}, is a newer decomposition technique that offers improved noise robustness and signal separation. Unlike empirical decomposition methods, VMD produces compact, non-overlapping modes with specific bandwidths. It has been successfully applied in various forecasting contexts but remains underexplored in cryptocurrency applications.\\

\noindent
This study proposes a hybrid forecasting model that integrates VMD with LSTM. The VMD component decomposes the raw Bitcoin price signal into several Intrinsic Mode Functions (IMFs), each of which is modeled by a separate LSTM network. The predictions are aggregated to form the final forecast. This structure is designed to better capture both the short- and long-term dynamics of Bitcoin price behavior.

\section{Related Work}

The volatility of Bitcoin has driven considerable interest in accurate forecasting methods. Early research explored statistical and probabilistic models. For example, Shah and Zhang \cite{shah2014bayesian} proposed a Bayesian regression model that yielded an 89\% return in 50 days with a Sharpe ratio of 4.1. Hegazy and Mumford \cite{hegazy2016comparitive} used decision-tree-based models leveraging price derivatives to predict price direction with 57.11\% accuracy.\\

\noindent
Classical time series models such as ARIMA have also been used for Bitcoin price forecasting. Karakoyun and Cibikdiken \cite{karakoyun2018comparison} compared ARIMA and Long Short-Term Memory (LSTM) networks, finding that LSTM achieved a significantly lower Mean Absolute Percentage Error (MAPE) of 1.40\% compared to 11.86\% for ARIMA.\\

\noindent
Subsequent studies explored advanced deep learning models. Awoke et al. \cite{awoke2020bitcoin} implemented both LSTM and GRU networks, concluding that GRU models offer better forecasting efficiency and shorter training times. Awotunde et al. \cite{awotunde2021machine} used recurrent neural networks (RNNs) and generalised regression neural networks (GRNNs) for multi-currency prediction. Atsalakis et al. \cite{atsalakis2019bitcoin} proposed a hybrid neuro-fuzzy controller that outperformed traditional models in exchange rate prediction.\\

\noindent
Several studies have demonstrated the strengths of machine and deep learning techniques such as XGBoost, AdaBoost, Transformers, LSTM, GRU, RNN models in modeling nonlinear, sequential dependencies that traditional baseline models tend not to capture \cite{nortey2025ai}. Ji et al. \cite{ji2019comparative} explored deep learning models such as DNNs, CNNs, and LSTM for Bitcoin price forecasting, while McNally et al. \cite{mcnally2018predicting} compared LSTM and RNN models with ARIMA and reported superior performance from deep learning methods.\\

\noindent
Despite these advances, many existing models struggle with nonstationarity and noise in financial time series \cite{nortey2024garch}. Buabeng et al. \cite{buabeng2022intelligent} argued that traditional models are inadequate for capturing complex nonlinear structures in Bitcoin price data. To address this, researchers have introduced signal decomposition techniques such as EMD and wavelets. However, mode mixing and redundancy remain major limitations.\\

\noindent
Variational Mode Decomposition (VMD), introduced by Dragomiretskiy and Zosso \cite{dragomiretskiy2013variational}, offers improved mode separation and noise robustness. It has shown promising results in fields such as wind forecasting and biomedical signal processing. However, its application in Bitcoin price prediction remains limited. This gap motivates the present study, which combines VMD with LSTM to construct a hybrid forecasting model capable of better capturing both low- and high-frequency components of the Bitcoin price series.

\section{Methodology}
This section outlines the methodological framework adopted in this study, which integrates Variational Mode Decomposition (VMD) and Long Short-Term Memory (LSTM) neural networks to forecast Bitcoin prices. The approach includes data acquisition and preprocessing, decomposition using VMD, LSTM model training for each decomposed component, and ensemble forecasting.

\subsection{Overview}
This study utilizes a hybrid forecasting approach combining VMD and LSTM. The hybrid model decomposes complex and non-stationary Bitcoin time series data into simpler components (Intrinsic Mode Functions or IMFs) using VMD, and then forecasts each component using a dedicated LSTM model. The final prediction is obtained by summing the forecasted components.

\subsection{Data}
The dataset consists of daily Bitcoin closing prices obtained from Yahoo Finance \cite{bitcoinprice} from August 11, 2017, to June 13, 2025. A total of 2,863 daily observations were recorded. After downloading, missing values were removed using the \texttt{dropna()} method. The 'Date' column was extracted from the index and converted to a regular column. An additional 'Year' column was created to support visual analysis. Summary statistics for the 'Close' price were computed, showing a minimum of 3,154.95 USD, a maximum of 111,673.28 USD, and a mean of 30,872.06 USD. The data was normalized using MinMax scaling.

\subsection{Variational Mode Decomposition (VMD)}
VMD is a technique that decomposes a time series signal into a finite number of band-limited intrinsic mode functions (IMFs), each with a narrow spectral band centered around an estimated center frequency. It solves the following constrained variational problem:

\begin{equation}
	\min_{\{u_k\}, \{\omega_k\}} \left\{ \sum_{k=1}^{K} \left\| \partial_t \left[ \left( \delta(t) + \frac{j}{\pi t} \right) * u_k(t) \right] e^{-j \omega_k t} \right\|_2^2 \right\}
\end{equation}

\begin{equation}
	\text{subject to } \sum_{k=1}^{K} u_k(t) = f(t)
\end{equation}
Here, $u_k(t)$ is the $k$-th IMF and $\omega_k$ is its center frequency. The number of modes $K$ was selected based on the residual energy threshold and set to $K = 15$.

\subsection{Long Short-Term Memory (LSTM) Network}

LSTM networks are a type of Recurrent Neural Network (RNN) designed to learn long-term dependencies in sequential data. Each LSTM unit contains a cell state and three gates: the forget gate, input gate, and output gate. The internal operations of an LSTM cell are defined as follows:

\begin{align}
	f_t &= \sigma(W_f [h_{t-1}, x_t] + b_f) \\
	i_t &= \sigma(W_i [h_{t-1}, x_t] + b_i) \\
	\tilde{c}_t &= \tanh(W_c [h_{t-1}, x_t] + b_c) \\
	c_t &= f_t \cdot c_{t-1} + i_t \cdot \tilde{c}_t \\
	o_t &= \sigma(W_o [h_{t-1}, x_t] + b_o) \\
	h_t &= o_t \cdot \tanh(c_t)
\end{align}

\noindent
Each IMF was normalized and passed to its own LSTM model. The models used a 30 time step lookback window. Training was performed using the Adam optimizer with a loss of mean squared error (MSE) over 20 epochs and a batch size of 32.

\subsection{Hybrid VMD-LSTM Forecasting Framework}
The VMD-LSTM forecasting process consists of the following phases:

\textbf{Decomposition:}
\begin{enumerate}
	\item Apply VMD with $K = 15$ to decompose the signal into $K$ IMFs.
	\item Evaluate residual energy to validate the quality of the decomposition.
\end{enumerate}

\textbf{Modeling:}
\begin{enumerate}
	\setcounter{enumi}{2}
	\item Generate supervised sequences using a 30-day lookback window for each IMF.
	\item Train a separate LSTM model for each IMF.
\end{enumerate}

\textbf{Forecasting:}
\begin{enumerate}
	\setcounter{enumi}{4}
	\item Predict the next 30 values for each IMF using its trained LSTM.
\end{enumerate}

\textbf{Ensemble:}
\begin{enumerate}
	\setcounter{enumi}{5}
	\item Combine the forecasts of all IMFs by summation:
	\begin{equation}
		\hat{y}(t) = \sum_{k=1}^{K} \hat{y}_k(t)
	\end{equation}
	
	\item Apply the inverse transformation to return to the original scale.
\end{enumerate}

\subsection{Model Evaluation}
The performance of each model was evaluated using standard regression metrics.Let $y_i$ be the actual value and $\hat{y}_i$ the predicted value.

\begin{align}
	\text{RMSE} &= \sqrt{\frac{1}{n} \sum_{i=1}^{n} (y_i - \hat{y}_i)^2} \\
	\text{MAE}  &= \frac{1}{n} \sum_{i=1}^{n} \left| y_i - \hat{y}_i \right| \\
	\text{MSE}  &= \frac{1}{n} \sum_{i=1}^{n} (y_i - \hat{y}_i)^2 \\
	R^2 &= 1 - \frac{\sum_{i=1}^{n} (y_i - \hat{y}_i)^2}{\sum_{i=1}^{n} (y_i - \bar{y})^2}
\end{align}

\noindent
The results were visualized and saved for further analysis.

\section{Results and Discussion}
\subsection{Overview}
This section presents the analysis and interpretation of the results obtained from the VMD-LSTM forecasting model. The discussion is structured around the model's descriptive analysis, decomposition performance, prediction accuracy, and comparison with a plain LSTM model.\\

\noindent
This study demonstrates that decomposing financial time series using VMD significantly enhances the forecasting capability of LSTM models. The hybrid VMD+LSTM model consistently outperformed the plain LSTM across all evaluation metrics and was also more effective in generating short-term forecasts. Future work can explore the integration of attention mechanisms or the application of VMD+LSTM to other financial instruments.

\subsection{Descriptive Analysis}
\autoref{tab:descriptive} shows the descriptive statistics of the daily Bitcoin closing prices between August 11, 2017 and June 13, 2025. The dataset comprises 2,863 observations. The average closing price during the period was approximately \$30,872.06, with a minimum of \$3,154.95 and a maximum of \$111,673.28. The distribution exhibits a right skew, as indicated by a median of \$22,487.39 which is less than the mean. The wide range and standard deviation of \$26,673.85 reflect high volatility in the Bitcoin market over the observation period.

\begin{table}[H]
	\centering
	\caption{Descriptive Statistics of Daily Bitcoin Closing Price}
	\label{tab:descriptive}
	\begin{tabular}{lr}
		\toprule
		Statistic & BTC-USD \\
		\midrule
		Count & 2,863 \\
		Mean & 30,872.06 \\
		Standard Deviation & 26,673.85 \\
		Minimum & 3,154.95 \\
		25\% Quantile & 8,803.47 \\
		Median (50\%) & 22,487.39 \\
		75\% Quantile & 46,352.58 \\
		Maximum & 111,673.28 \\
		\bottomrule
	\end{tabular}
\end{table}

\subsection{Residual Energy and Mode Selection (K)}
Before performing Variational Mode Decomposition (VMD), the number of modes $K$ must be carefully selected. This is critical, as an inappropriate value for $K$ can lead to under-decomposition or over-decomposition. If $K$ is too small, the signal is not adequately decomposed, potentially retaining noise and non-stationary characteristics. Conversely, a very large $K$ may over-segment the signal, introducing redundancy and increasing computational complexity without meaningful gains in accuracy.\\

\begin{figure}[H]
	\centering
	\includegraphics[width=0.75\textwidth]{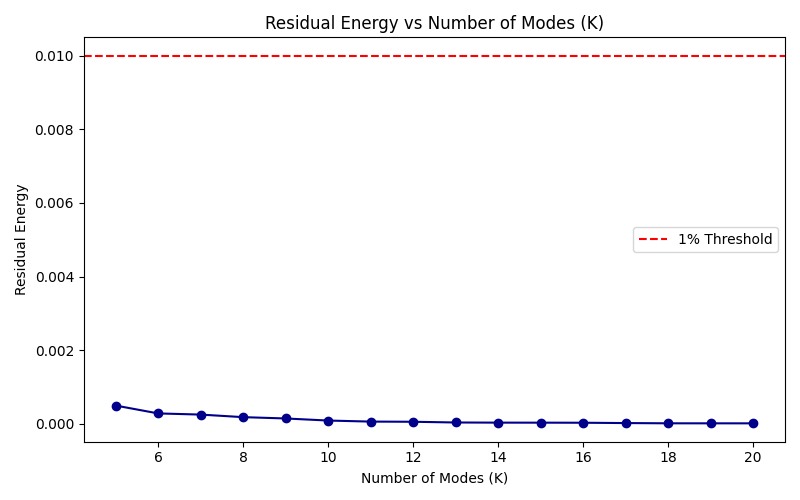}
	\caption{Residual Energy vs. Number of Modes (K)}
	\label{fig:residual_energy_plot}
\end{figure}

\noindent
To determine the optimal number of modes, the residual energy ratio was computed for values of $K$ ranging from 5 to 20.   \autoref{fig:residual_energy_plot} show the results. The residual energy consistently decreased as $K$ increased, with diminishing returns observed beyond $K = 10$. For $K \geq 11$, the residual energy dropped below 0.0001 and plateaued with negligible improvement.\\

\noindent
Therefore, $K = 15$ was selected for this study. This choice represents a balance between sufficient decomposition and computational efficiency. It ensures that essential signal components are preserved while minimizing noise, thus improving forecasting performance.

\subsection{IMF Decomposition Results}
Following the selection of $K = 15$ based on the residual energy criterion (see  \autoref{fig:residual_energy_plot}), the Variational Mode Decomposition (VMD) algorithm was applied to the normalized Bitcoin closing price series. This process resulted in 15 Intrinsic Mode Functions (IMFs), each representing a distinct frequency band of the original signal.\\

\begin{figure}[H]
	\centering
	\includegraphics[width=\textwidth]{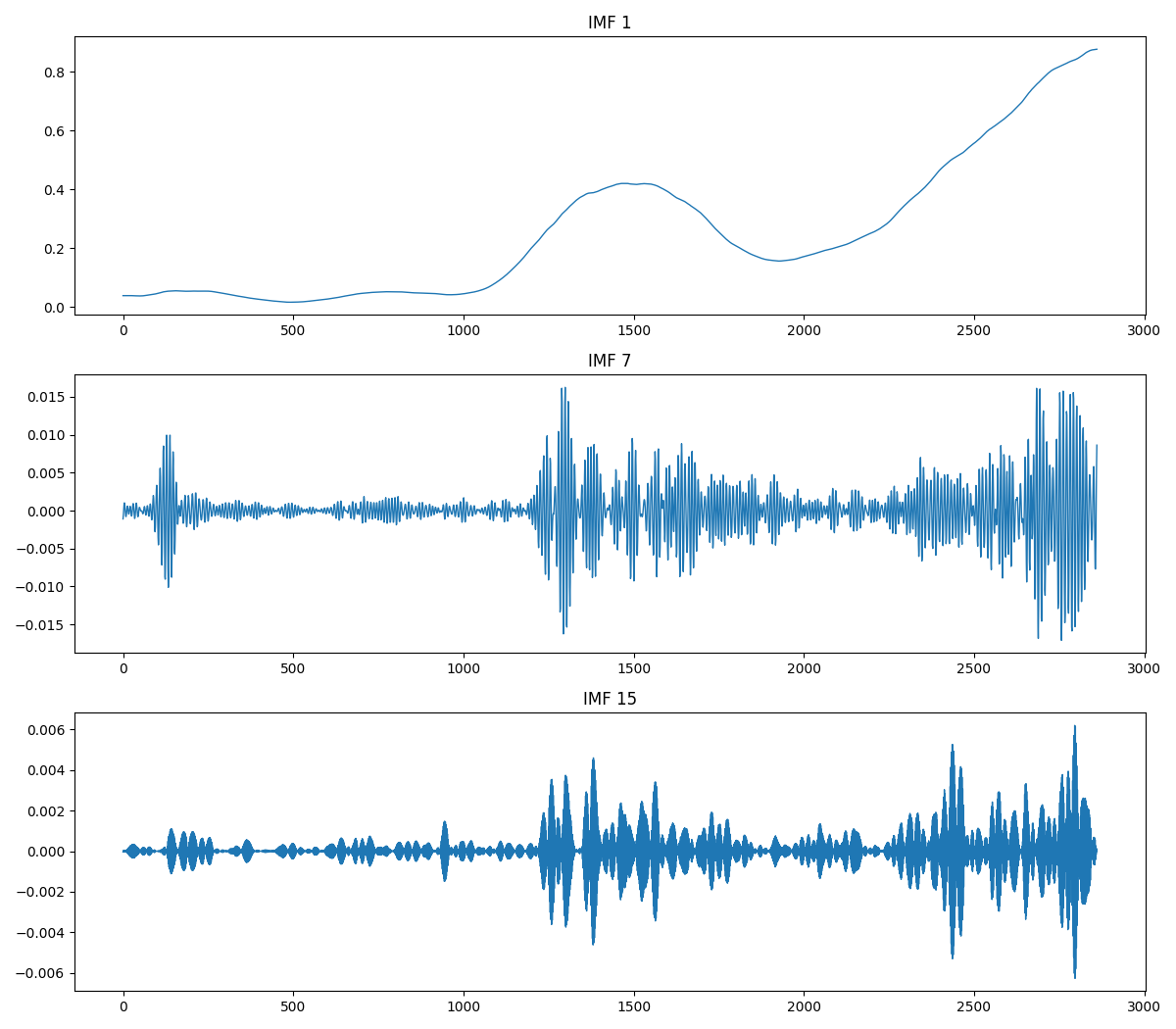}
	\caption{Selected IMF Components from VMD Decomposition of Bitcoin Price}
	\label{fig:imf_decomposition}
\end{figure}

\noindent
 \autoref{fig:imf_decomposition} shows three representative Intrinsic Mode Functions (IMFs) obtained from the VMD decomposition of the normalized Bitcoin closing price series.

\begin{itemize}
	\item \textbf{IMF 1} captures the long-term trend of the Bitcoin price. It reflects the overall movement and structural pattern of the series across the full time horizon.
	
	\item \textbf{IMF 7} exhibits medium-frequency oscillations, which may correspond to market cycles or seasonal fluctuations. These components reveal intermediate-term patterns that are neither too smooth nor too volatile.
	
	\item \textbf{IMF 15} contains high-frequency and low-amplitude fluctuations, often associated with short-term noise or abrupt market changes. It captures the fine-grained variations in the data.
\end{itemize}

\noindent
The use of VMD allows the complex, non-stationary Bitcoin price series to be broken down into distinct frequency components, each capturing different aspects of the signal. By modeling these components separately, the hybrid approach can better account for both long-term trends and short-term dynamics. This not only improves the interpretability of the data but also contributes to more accurate and robust forecasting performance.

\subsection{VMD+LSTM Model}
\begin{figure}[H]
	\centering
	\includegraphics[width=0.6\textwidth]{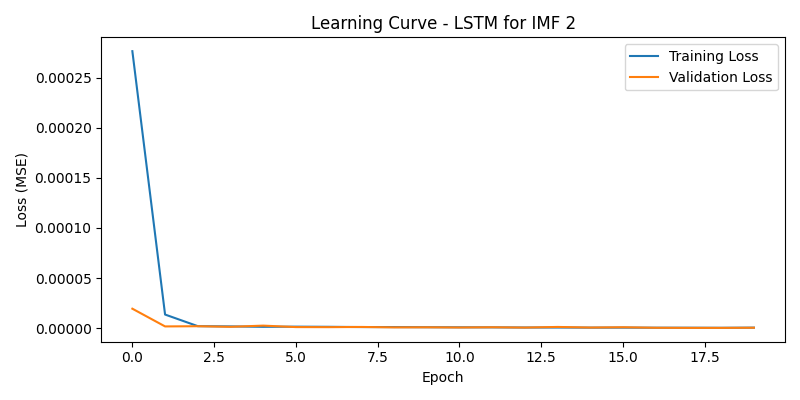}
	\includegraphics[width=0.6\textwidth]{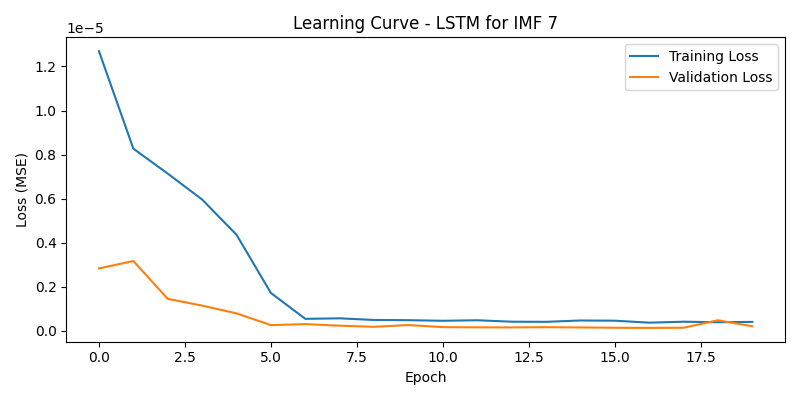}
	\includegraphics[width=0.6\textwidth]{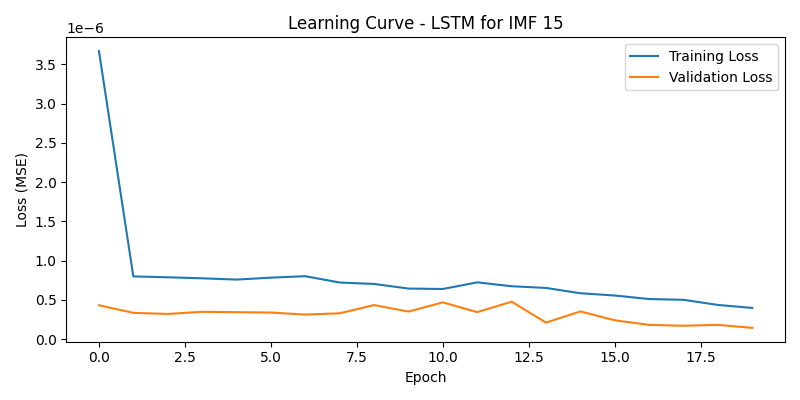}
	\caption{Learning Curves for IMF 2, 7, and 15}
	\label{fig:learning_curves_imf}
\end{figure}

The learning curves for selected IMFs (2, 7, and 15) are shown in \autoref{fig:learning_curves_imf}.The curves demonstrate a consistent reduction in training and validation losses, suggesting an effective model fitting without overfitting.

\begin{figure}[H]
	\centering
	\includegraphics[width=0.95\textwidth]{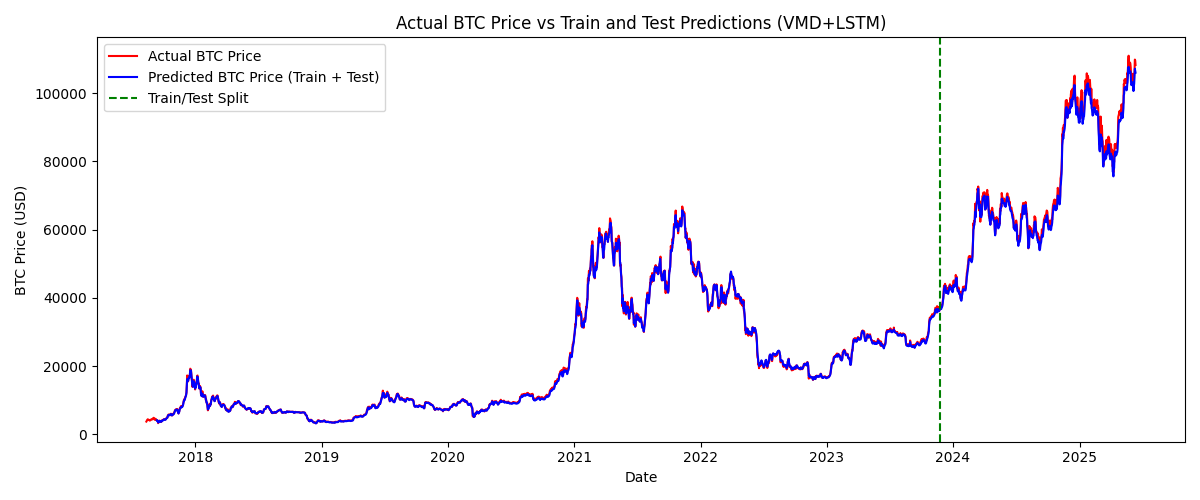}
	\caption{Actual vs. Predicted Bitcoin Prices (VMD+LSTM)}
	\label{fig:btc_actual_vs_pred}
\end{figure}

\noindent
\autoref{fig:btc_actual_vs_pred} illustrates the actual Bitcoin closing prices (in red) compared with the predicted prices (in blue) generated by the VMD+LSTM model across both the training and testing phases. The green vertical dashed line represents the train-test split. The close alignment between the actual and predicted prices in both segments highlights the model’s ability to accurately capture the nonlinear patterns and long-term dependencies in the time series data. The model performed particularly well during periods of high volatility, demonstrating robustness in predicting extreme price movements and validating the effectiveness of the hybrid VMD+LSTM forecasting framework.\\

\begin{figure}[H]
	\centering
	\includegraphics[width=0.85\textwidth]{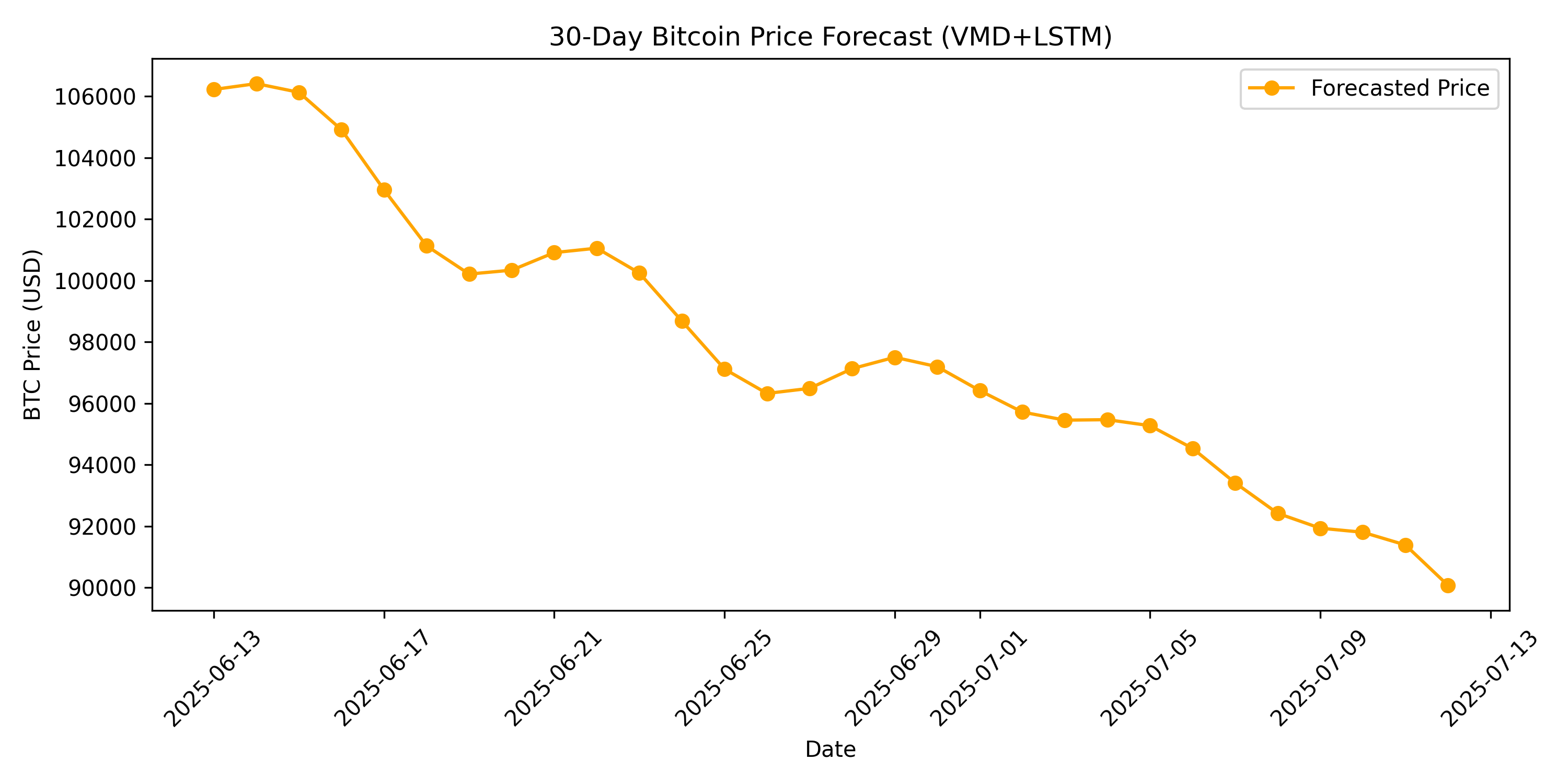}
	\caption{30-Day Bitcoin Price Forecast using VMD-LSTM}
	\label{fig:forecast_30days}
\end{figure}

\noindent
\autoref{fig:forecast_30days} illustrates the 30-day Bitcoin price forecast using the proposed VMD-LSTM model. The forecast spans from June 13, 2025, to July 12, 2025. As observed from the plot, the predicted trend captures a gradual decline in Bitcoin price over the forecast window, with minor fluctuations around short-term levels. This indicates that the model has learned meaningful temporal dependencies from the decomposed IMFs and is capable of generating a stable forecast trajectory.\\

\noindent
The downward pattern shown in the prediction aligns with recent trends in the tail end of the training data, suggesting that the model successfully captured the momentum shift. The smoothness and consistency of the forecasted values, without abrupt spikes, further highlight the model’s ability to reduce noise and volatility in the input signal — one of the key motivations for incorporating VMD.

\subsection{LSTM Model}
\begin{figure}[H]
	\centering
	\includegraphics[width=0.85\textwidth]{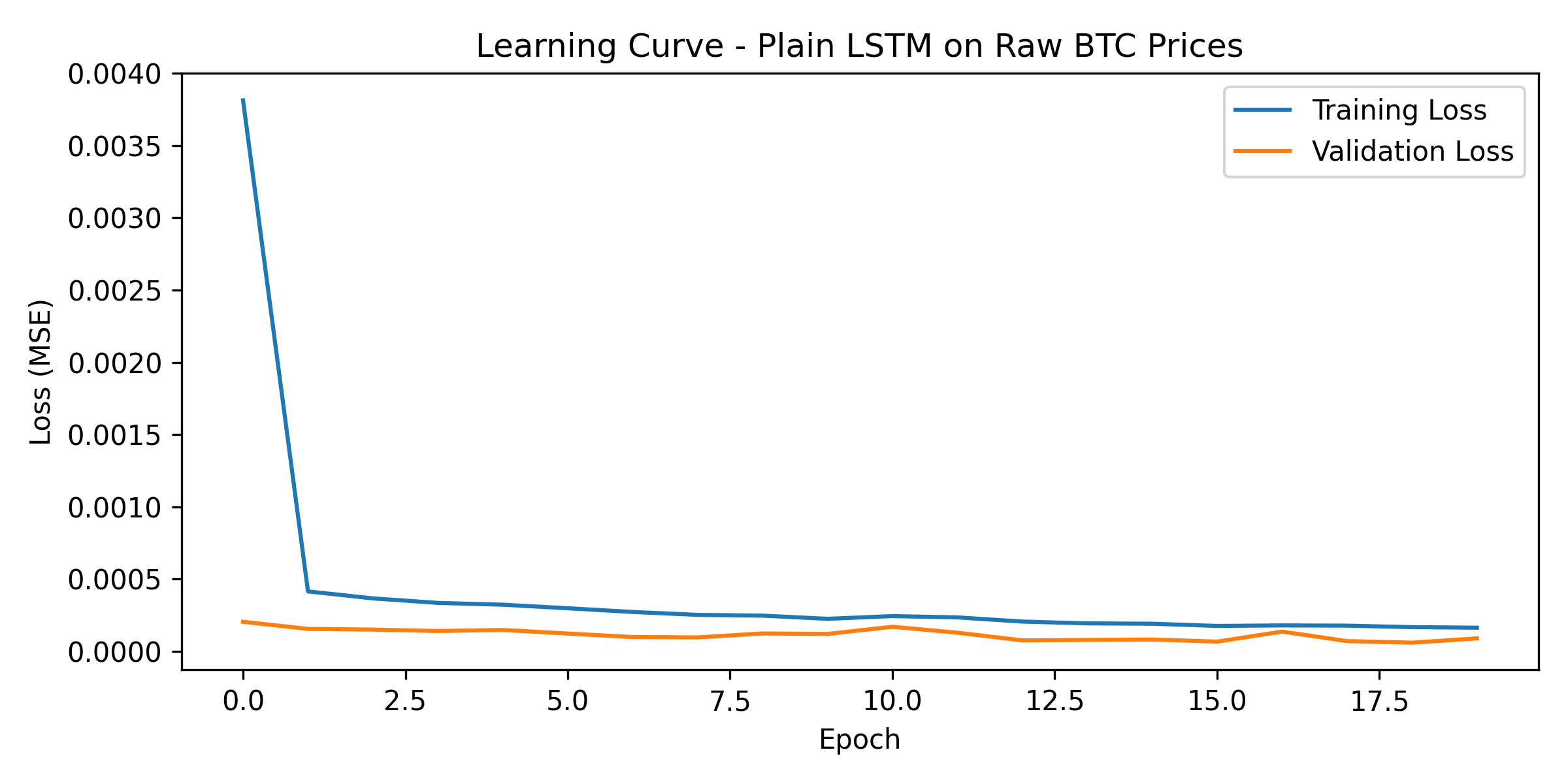}
	\caption{Learning Curve of Plain LSTM Model on Raw Bitcoin Prices.}
	\label{fig:plain_lstm_learning_curve}
\end{figure}

\autoref{fig:plain_lstm_learning_curve} shows the training and validation loss curves for the plain LSTM model applied directly to the raw Bitcoin price data. Both curves decline consistently, indicating that the model learned the patterns in the training data without signs of overfitting. However, even with this apparent convergence, the final loss values and evaluation metrics (RMSE, MAE, MSE, and $R^2$) remained less favorable compared to those from the VMD-enhanced LSTM model. This result highlights the importance of using VMD to decompose and denoise the original signal, which helps improve forecasting accuracy.

\subsection{Comaparison of VMD+LSTM and Plain LSTM}
To evaluate the effectiveness of the proposed hybrid model, the performance of the VMD+LSTM model was compared against a plain LSTM model trained directly on the original (non-decomposed) Bitcoin closing prices.  \autoref{tab:comparison} presents a comparative analysis using four standard evaluation metrics: RMSE, MAE, MSE, and $R^2$.\\

\noindent
The results show that the VMD+LSTM model outperformed the plain LSTM on both training and test sets. Notably, the RMSE and MAE values for the VMD+LSTM model on the test set were significantly lower (RMSE: 1619.63 vs. 4005.32; MAE: 1397.18 vs. 3157.91), indicating more accurate predictions. Similarly, the mean squared error (MSE) dropped from over 16 million in the plain LSTM to approximately 2.6 million with VMD+LSTM, demonstrating reduced prediction error.\\

\noindent
In addition, the $R^2$ values — which measure the proportion of variance explained by the model — remained very high for both models but were slightly better in the VMD+LSTM (0.9995 train / 0.9934 test) compared to the plain LSTM (0.9922 train / 0.9598 test). This suggests that the hybrid approach captured the underlying patterns in the data more effectively.\\

\noindent
Overall, the decomposition of the original signal into smoother Intrinsic Mode Functions (IMFs) before applying LSTM not only enhanced predictive performance but also reduced overfitting, as evidenced by the lower test error.
\begin{table}[H]
	\centering
	\caption{Comparative Analysis of Hybrid VMD-LSTM Model and Plain LSTM}
	\label{tab:comparison}
	\begin{tabular}{lcccc}
		\toprule
		\textbf{Metric} & \multicolumn{2}{c}{\textbf{VMD-LSTM}} & \multicolumn{2}{c}{\textbf{Plain LSTM}} \\
		\cmidrule(lr){2-3} \cmidrule(lr){4-5}
		& \textbf{Train} & \textbf{Test} & \textbf{Train} & \textbf{Test} \\
		\midrule
		RMSE & 352.8744 & 1619.6331 & 1398.54 & 4005.32 \\
		MAE  & 240.3882 & 1397.1762 & 918.58  & 3157.91 \\
		MSE  & 124520.3444 & 2623211.3442 & 1955913.88 & 16042578.17 \\
		$R^2$ & 0.9995 & 0.9934 & 0.9922 & 0.9598 \\
		\bottomrule
	\end{tabular}
\end{table}
\section{Conclusion}
This study set out to develop a hybrid deep learning forecasting model that combines Variational Mode Decomposition (VMD) with a Long Short-Term Memory (LSTM) network to forecast Bitcoin prices. Additionally, a comparative analysis was conducted against a standard LSTM model trained directly on the raw data. The results demonstrate that both models are capable of forecasting Bitcoin prices, with the VMD-LSTM model achieving superior performance. Specifically, the VMD-LSTM model attained an impressive $R^2$ of 0.9934, compared to 0.9598 for the standard LSTM. Furthermore, the hybrid model consistently outperformed the standard LSTM in all evaluation metrics, including lower RMSE, MAE, and MSE, and a higher coefficient of determination ($R^2$).\\

\noindent
These findings suggest that the VMD-based preprocessing step effectively reduces the non-stationarity and complexity in the time series, leading to improved forecasting accuracy. The VMD-LSTM approach, therefore, holds significant promise for applications in cryptocurrency forecasting and can be extended to other domains involving complex financial or medical time series data.

\subsection*{Declaration of competing interest}
\noindent
The authors declare that they have no known competing financial interests or personal relationships that could have appeared to
influence the work reported in this paper.

\subsection*{CRediT author statement}
The submitted version of the article was approved by all authors who contributed equally.

\subsection*{Data Availability}
\noindent
The data used to support the findings of this study are available upon reasonable request from the corresponding author.

\subsection*{Acknowledgment}
\noindent
The first and corresponding author also acknowledges the enormous support of the University of Texas Rio Grande Valley (UTRGV).

\bibliographystyle{unsrt} 
\bibliography{REFERENCES} 
\end{document}